\pdfoutput=1
\documentclass[11pt]{article}
\usepackage{latexsym}

\usepackage{amssymb}
\usepackage{amsmath}

\usepackage{epsfig}
\usepackage{graphicx}

\numberwithin{equation}{section}  

\newcommand{\be}{\begin{equation}}
\newcommand{\ee}{\end{equation}}
\newcommand{\ba}{\begin{eqnarray}}
\newcommand{\ea}{\end{eqnarray}}
\newcommand{\baa}{\begin{array}}
\newcommand{\eaa}{\end{array}}
\newcommand{\bi}{\begin{itemize}}
\newcommand{\ei}{\end{itemize}}
\newcommand{\edoc}{\end{document}}

\newcommand{\nn}{\nonumber \\}
\newcommand{\nr}[1]{(\ref{#1})}
\newcommand{\la}[1]{\label{#1}}

\newcommand{\fra}[2]{{\textstyle{\frac{#1}{#2}\,}}}  
\newcommand{\mn}{{\mu\nu}}

\newcommand{\zb}{{\bar z}}

\newcommand{\cb}{{\bar c}}

\newcommand{\p}{{\partial}}

\def\Tr{{\rm Tr\,}}

\def\CI{{\cal I}}
\def\gsim{\raise0.3ex\hbox{$>$\kern-0.75em\raise-1.1ex\hbox{$\sim$}}}
\def\lsim{\raise0.3ex\hbox{$<$\kern-0.75em\raise-1.1ex\hbox{$\sim$}}}

\begin{document}

\begin{titlepage}
\begin{flushright}
HIP-2019-09/TH\\
\end{flushright}
\begin{centering}
\vfill

{\Large{\bf Memory effect in Yang-Mills theory}}

\vspace{0.8cm}

\renewcommand{\thefootnote}{\fnsymbol{footnote}}

Niko Jokela$^{1,2,*}$, K. Kajantie$^{2,\dagger}$, Miika Sarkkinen$^{1,\ddagger}$

\vspace{0.8cm}
$^1$Department of Physics and $^2$Helsinki Institute of Physics,\\ FI-00014 University of Helsinki, Finland
 
 \vspace{0.8cm}
\end{centering}

 \vspace{0.8cm}
\noindent
We study the empirical realisation of the memory effect in Yang-Mills theory, especially in view of the classical vs. quantum nature of the theory.  Gauge invariant analysis of memory in classical U(1) electrodynamics and its observation by total change of transverse momentum of a charge is reviewed. Gauge fixing leads to a determination of a gauge transformation at infinity. An example of Yang-Mills memory then is obtained by reinterpreting known results on interactions of a quark and a large high energy nucleus in the theory of Color Glass Condensate. The memory signal is again a kick in transverse momentum, but it is only obtained in quantum theory after fixing the gauge, after summing over an ensemble of classical processes.

\vfill

\hrule{}
\vspace{0.4cm}
\hspace{-4mm}$*$ niko.jokela@helsinki.fi\\
$\dagger$ keijo.kajantie@helsinki.fi\\
$\ddagger$ miika.sarkkinen@helsinki.fi
\end{titlepage}

\section{Introduction}
The memory effect in gravitational radiation \cite{ww2,garfinklegrav} is the total change in the positions (or other properties)
of a system of detectors left by a burst of gravitational radiation. Conceptually, the detectors lie
at null infinity and that is where massless gravitons end up. The effect can also be formulated for other massless 
(or nearly massless) particles like photons \cite{bieriED,winicour} or neutrinos \cite{bieri2}.  
Massless quanta exist also in Yang-Mills (YM) theories and color memory has been
studied in \cite{pate} in a classical perturbative approximation, for which 
the discussion of electrodynamics \cite{stromED} can directly be extended.

Yang-Mills fields form an integral part of the standard model (SM), SU(3) invariance in the strong quark-gluon sector and
SU(2)$\times$U(1) invariance in the electroweak sector. However, in both cases the theory is in a phase 
which, apart from the photon, does not contain massless particles to be sent to null infinity. In fact, this is the very reason for their being the
SM. The SU(3) sector is in the confinement phase,  massless gluons do not propagate but
develop a gap, become massive glueballs and do not go to null infinity. The SU(2)$\times$U(1) sector is in
a Higgs phase, the candidate gluons develop mass via the Higgs mechanism and, after mixing with U(1), become
three massive vector bosons and the single massless photon. For a lucid comparative exposition of the SM confinement and
Higgs phases, see \cite{thooft}.

There is one phenomenological context in which classical Yang-Mills fields are studied with some
justification: wave function of a heavy nucleus when probed with a large scale probe like an
electron in deep inelastic scattering  \cite{iancu}. A large nucleus probed with large $Q^2$ involves large
occupation numbers and hence classical fields. The realisation of the ideas in \cite{pate,stromED} in this
context has already been studied in \cite{ymmemo2}. We wish in this note to give a 
simple discussion of YM memory in the spirit of \cite{bieriED,winicour}, emphasising the fact that
any experimental measurement of the suggested YM effect is inherently quantum mechanical. Here the enormous
non-linear complexity of classical theory is replaced by a relative simplicity of quantum mechanical expectation values.
As a background we first discuss the memory in electrodynamics and explain why we feel that the ``new
symmetries of QED'' \cite{stromED} are basically U(1) invariance of classical ED, when 
applied in a fixed gauge at null infinity.

Actually the place where SU(3) YM memory effect empirically appears 
is very easy to locate and well known in the field, just nomenclature has to be changed.
For example, Fig. 12 of \cite{iancu} shows how a passage of a large energy nucleus creates from vacuum a transverse matrix 
color field $A^i,\,\,i=2,3$, also gauge equivalent to vacuum, $F_{ij}=0$. The nucleus is the analogue of a burst of
YM radiation, the color field $A^i$ is the analogue of its memory or of the ``large'' gauge transformation
in \cite{pate}. More precisely, $A^i$ is the (square root of) the quantum expectation value of the square of the
YM field, summed over colors. Quantum physics cannot be avoided.

A basic memory signal will be transverse momentum kick of a test object. In U(1) one is hereby done, no problem in measuring
the kick. In YM one further has to see how this kick is measured. These measurements are rather indirect, as thoroughly discussed 
in \cite{ymmemo2}. Another possibility would be to use the
gluon radiation from the acceleration of the test quark, as recently computed in \cite{Kajantie:2019hft}. The use of YM memory as
an analogue of gravitational wave memory is thus very limited.

The electroweak sector of the SM can also be forced out of the Higgs phase by similar means, by large occupation numbers and
associated classical fields in the very early universe. The boson equilibrium occupation number at small $k$ is $n=1/(e^{k/T}-1)\sim
T/k\sim 1/g^2\gg1$ since the dominant infrared scale is the coupling of the 3d magnetic sector of the theory, $k\sim g^2T$.
Important physical phenomena like baryon number violation rates in the SM can be numerically
studied in this setting, see for example \cite{rummu1,rummu2}. A study of YM memory also for these fields
should be possible, also in an expanding universe  \cite{riotto,chu}.

In discussions of memory effect one usually thinks about radiation propagating over huge distances, millions of light
years. In the heavy ion Yang-Mills case one clearly must be content with much smaller distances. This is related to
the coupling not being asymptotically small at the relevant scale of about 1 GeV.
Phenomenologically the coupling constant $g(\mu=1\,{\rm GeV})\approx2$ is actually ``large'' 
in the usual $\overline{\rm MS}$ renormalisation
scheme in the sense that the distance scale generated by renormalisation is ``small''. For $N_c=N_f=3$,
\be
{1\over\Lambda_{\rm QCD}}={1\over\mu} \exp\left[{8\pi^2\over 9g^2(\mu)}\right]\biggl({9g^2\over 16\pi^2}\biggr)^{{32\over 81}}
\approx 1\,{\rm fm} \ .
\ee
Theoretically, of course one can
apply much smaller values of $g^2$ and there is underway an intense numerical effort for studying classical YM equations
in the weak coupling region
(see, for example, \cite{kurkela1,kurkela2}). 
In the SM, the corresponding SU(2) coupling constant $g(\mu=100\, {\rm GeV})\approx2/3$ is
actually small in the sense that the corresponding distance scale is macroscopic. Including just the
2-loop running of the SU(2) coupling \cite{steinhauser}, the renormalisation group integration constant is
\be
{1\over\Lambda_2}={1\over\mu}\exp\left[{48\pi^2\over 19g^2(\mu)}\right]\biggl({19g^2\over 96\pi^2}\biggr)^{{483\over 361}}\ .
\ee
Putting here $g(\mu=100\,{\rm GeV})=\fra23$ gives an SU(2) distance scale of about $1/\Lambda_2= 4000$ km, a 
macroscopic distance. However, 
the standard model is in the Higgs phase so that this is not a proper confinement radius within which the fields would be massless.

In the following we shall
limit ourselves to a summary of the memory in U(1) ED (Section 2) and a discussion of memory in SU(3) heavy ion
collisions (Sections 3 and 4). Note on angular coordinates at large distances: for ED we use two angular coordinates
$\theta_A=h_{AB}\theta^B$ on S$^2$, while for the heavy ion case we use two Cartesian coordinates $x_i=x^i,\, i=2,3$ transverse to the
beam direction $x^1$. We use the mostly plus metric, its advantage is that
one can write $x_i=x^i$ without sign change.

\section{Memory in electrodynamics}
For ED everything follows from Maxwell's equations, evaluated at $\CI^+$, future null infinity. 
The essence of the phenomenon can be summarised as follows, in the spirit of \cite{bieriED}.

We use the coordinates $t,r,\theta^A$ and the metric
\be
g_\mn=\left( \begin{array}{ccc}-1 & 0 & 0  \\ 0 & 1 & 0 \\ 0&0& r^2h_{AB} \end{array} \right),\quad
g^\mn=\left( \begin{array}{ccc}-1 & 0 & 0  \\ 0 & 1 & 0 \\ 0&0& h^{AB}/r^2 \end{array} \right)\ ,
\la{eflat}
\ee
where $h_{AB}$ is the metric on the celestial sphere S$^2$, with, for example, $\theta^A=(\theta,\phi)$ in standard
spherical coordinates or $\theta^A=(z=e^{i\phi}/\tan(\theta/2),\zb)$ in stereographic coordinates. 
Maxwell's equations are 
\be
\nabla_\mu F^\mn = {1\over\sqrt{-g}}\partial_\mu(\sqrt{-g}F^\mn)=J^\nu,\qquad 
\la{max1}
\ee
\be
\nabla_\alpha F_{\beta\gamma}+\nabla_{\beta} F_{\gamma\alpha}+\nabla_\gamma F_{\alpha\beta}=
\partial_\alpha F_{\beta\gamma}+\partial_{\beta} F_{\gamma\alpha}+\partial_\gamma F_{\alpha\beta}=0 \ .
\la{max2}
\ee
Splitting vectors in their radial and celestial sphere S$^2$ components, $E^a=(E^r, E^A)$,
$E^A=h^{AB}E_B/r^2$, 
the inhomogeneous  \nr{max1} and homogeneous equations \nr{max2} are
\ba
J^t&=&\fra1{r^2}\p_r(r^2E_r)+D_AE^A,\nn
J^r&=&-\p_t E_r+\epsilon^{AB}D_AB_B,\la{inhomeqs}\\
J_A&=&-\p_tE_A+\epsilon_A^{\hspace{2mm}B}(\p_B B^r-\p_rB_B).\nonumber\\
0&=&\fra1{r^2}\p_r(r^2B_r)+D_AB^A,\nn
0&=&\p_t B^r+\epsilon^{AB}D_BE_A,\la{rcomplt}\\
0&=&\p_tB_A-\epsilon_A^{\hspace{2mm}B}(\p_B E_r-\p_r E_B),\nonumber
\ea
where the 2d epsilon tensor is ($h=\det h_{AB}$)
\be
\epsilon_{ab}=\sqrt{|h|}\left( \begin{array}{cc}0 & 1 \\ -1 &  0 \end{array} \right)\equiv \sqrt{|h|}\,\eta_{ab}.
\la{2deps}
\quad
\epsilon^{ab}={1\over h}\epsilon_{ab}={{\rm sign}\,\,h\over\sqrt{|h|}}\left( \begin{array}{cc}0 & 1 \\ -1 &  0 \end{array} \right)
={{\rm sign}\,\,h\over\sqrt{|h|}}\eta_{ab} \ .
\ee

\begin{figure}[!t]
\begin{center}

\vspace{-1.8cm}

\includegraphics[width=0.6\textwidth]{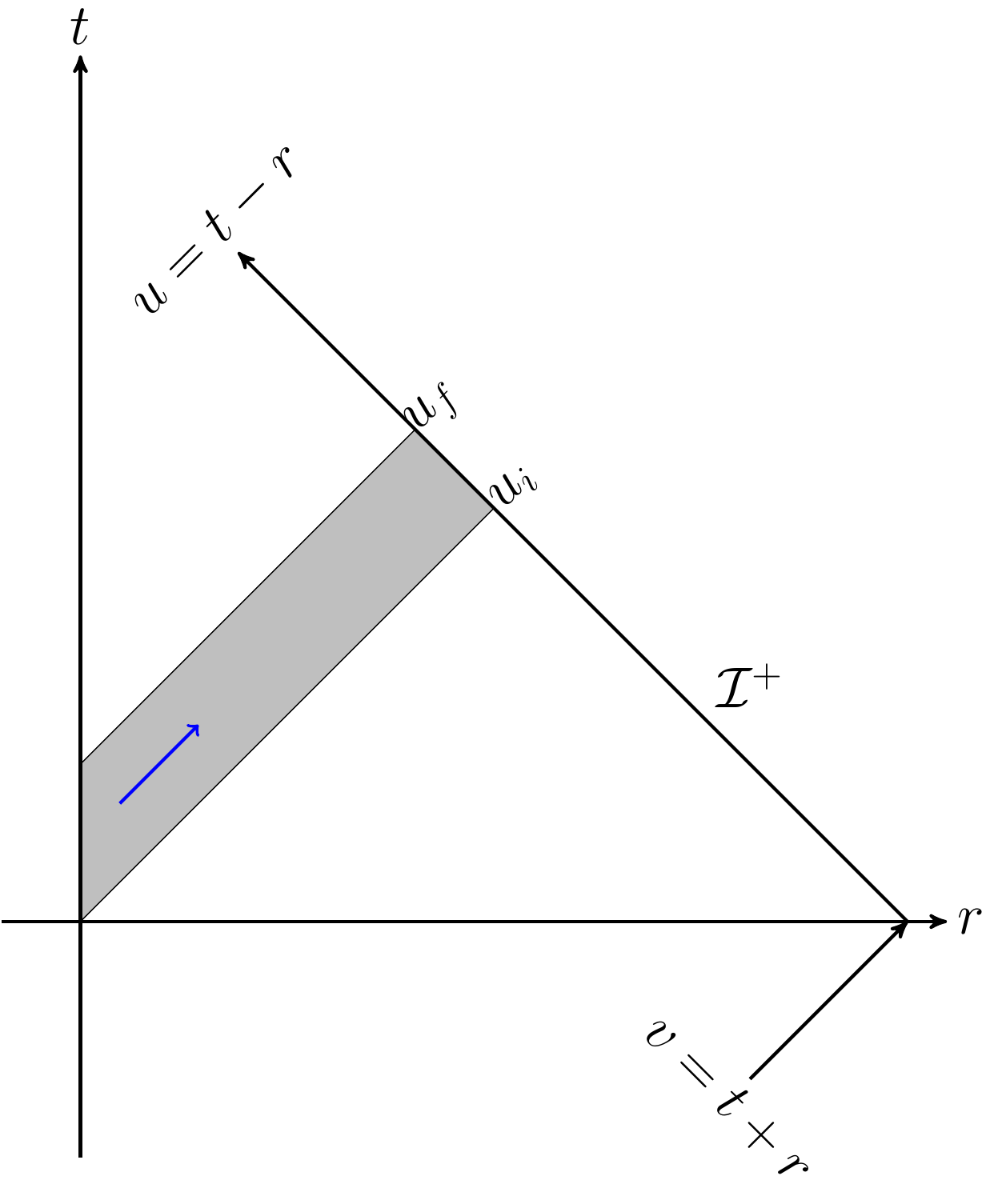}

\end{center}

\vspace{-1.0cm}
\caption{\small  Memory effect in electrodynamics. A radiator at $r=0$ sends a pulse of radiation to
null infinity ${\cal I}^+$ during the time interval $u_i<u<u_f$. The time integrated pulse of transverse electric field
gives a total momentum kick in \protect\nr{memory} to a test charge at null infinity.
}
\la{burstED}
\end{figure}

To analyse the memory we need a radial current at null infinity, i.e., 
\be
J^t=\rho={L(u,\theta^A)\over r^2}=J^r,\quad u=t-r\ .
\ee
The $r$ dependence is constrained by the total luminosity $\sim L(u,\theta^A)$ being finite.
The transverse current can be negligible, i.e., $J_A\sim 1/r^3$. Note that this current pattern
rigorously speaking  
implies that we should have massless charged particles, since only they get to null
infinity. This is a practical issue, though. For observations null infinity is at finite distance.

We now expect at null infinity a typical radiation pattern with transverse and orthogonal
electric and magnetic fields. To analyse the magnitudes it is enough to focus on the Gauss'
law, the first equation in \nr{inhomeqs}. 
Canceling a common factor $1/r^2$, replacing $\p_r\to -\p_u$ (at $\CI^+$ fields are functions of $t-r$)
and letting $E_r$ denote the leading $r$ behaviour, $E_r^{(2)}=r^2E_r\to E_r$ the equation is
\be
-\p_u E_r+D_AE^A=L\ .
\ee
Integrating this over $u_i<u<u_f$ gives
\be
E_r(u_i)-E_r(u_f)+D_A\int du\, E^A = \int du\, L(u,\theta^A)\equiv F(\theta^A)\ ,
\ee
which we write in the form
\be
D_AM^A=\Delta E_r + F(\theta^A), \la{fund}
\ee
defining the fundamental quantity, the memory vector
\be
M_A(\theta^A)=\int_{u_i}^{u_f} du\,E_A\ .
\la{MA}
\ee
The memory vector represents the cumulative effect on the celestial sphere of a pulse of radiation 
sent to $\CI^+$. Its significance is based on the fact that it is measurable. Integrating the
Lorentz force equation
\be
{dp^\mu\over d\tau}=q F^\mn u_\nu,\quad p^\mu=mu^\mu\ ,
\ee
for $\mu=B$ and for small velocities over time (effectively the same as $u$)
the change of the transverse momentum is
\be
\int dt {dp_B\over dt}=\Delta p_B = q\int dt\, E_B(t) = q M_B(\theta^A)\ .
\la{memory}
\ee
The pulse changes the momentum of a test particle by an amount given by the memory
vector.  This transverse kick is the simplest version of the electromagnetic memory effect.
It has two parts, an ordinary kick due to the change of the radial component of the 
electric field and a null kick due to flux of charge to null infinity, see Fig.~\ref{burstED}.

Integrating \nr{fund} over the celestial sphere, defining $\langle O\rangle = \int d\Omega \,O$,
one has
\be
\langle D_AM^A \rangle = 0 = \langle E_r(u_f)\rangle -\langle E_r(u_i)\rangle + \langle F \rangle = 
Q_f-Q_i+\langle F \rangle\ .
\ee
Here the first step is the fact that an integral of the divergence of a vector field over S$^2$
vanishes. Further one uses Gauss law, the integral of $r^2E_r$ gives the charge inside the
sphere. The equation thus expresses the fact that the $L(u)$ term has carried through
the sphere the amount $Q_i-Q_f$ of charge.

Given the charge density $L(u,\theta^A)$ at $\CI^+$
and the change in the radial electric field (here a standard example is a charged particle
initially at rest and then moving with constant velocity) one can compute the memory vector,
as concretely discussed in \cite{bieriED}.

Note that the above discussion is entirely covariant and in terms of fields, vector potentials
with some gauge choice have not been used. Consider, however, what happens
if one uses the coordinates $u,r,\theta^A$ and chooses the temporal gauge $A_u=0$
(equivalently, one could choose $A_r=0$ and further $A_u=0$ at 
a fixed value of $r$, $r=\infty$, at $\CI^+$) \cite{winicour}.
Then $E_A=F_{uA}=\p_u A_A-\p_A A_u=\p_u A_A$ and the memory vector and the associated
kick are, from \nr{memory}
\be
\Delta p_B=qM_B=q\int du\,E_B =q\int du\,\p_uA_B=q(A_B(u_f,\theta^A)-A_B(u_i,\theta^A))\ .
\la{MADeltaA}
\ee
We have thus used a physical measurement to determine a gauge choice dependent
quantity.  Nothing has happened to the symmetry properties of electrodynamics, it is
still U(1) gauge invariant. Interpreted as asymptotic symmetries at null infinity these
can be described as new symmetries \cite{stromED}, but physically there is nothing
beyond U(1) gauge invariance.

In the literature there are no suggestions of how to realize the ED kick memory in an experimental setup. However, even at a theoretical level it serves to elucidate some aspects of gravitational radiation memory that have remained unclear until recent years. In particular, it serves to underline the fact that there is a distinction between memory due to sources that do not get to null infinity and sources that do, or between ordinary and null memories respectively. Therefore, it may be useful to outline the relation of the above to gravitational radiation
and its memory effect. What is the ``gauge invariance'', how is the gauge ``fixed''  and what are the ``gauge
transformations'' the parameters of which are determined by measuring the memory effect?

Gauge invariance obviously is the diffeomorphism invariance of general relativity and gauge fixing is
finding the metric containing gravitational radiation, exactly as the Schwarzschild metric
contains a black hole. Using the coordinates $u,r,z,\zb$, where $z,\zb$ are the standard stereographic
coordinates on the celestial sphere S$^2$, this is the Bondi metric \cite{stromingerlectures}
\be
g_\mn=\left( \begin{array}{cccc}-1+{2Gm(u)\over r} & -1 & u_z(z,\zb,u) & u_\zb(z,\zb,u)  \\ -1 & 0 & 0 &0 \\ 
u_z(z,\zb,u) & 0 & r c(z,\zb,u) & r^2 \gamma \\ u_\zb(z,\zb,u) & 0 &r^2 \gamma  & r {\bar c}(z,\zb,u)  \end{array} \right),
\quad \gamma={2\over (1+z \zb)^2},\quad u_z={1\over 2\gamma}\p_\zb c,
\la{bondim}
\ee
$u_\zb$ is defined similarly in terms of $\cb$. 
This metric is defined near null infinity, $u=$ constant, $r\to\infty$ and contains subtle large $r$ corrections to the
flat metric,  a Schwarzschild-metric--like but time dependent mass term (which leads to the Vaidya
model) and a $1/r$ correction to the S$^2$ metric, specified by the functions $c(z,\zb,u),\,\cb(z,\zb,u)$. 
Time dependence of the mass represents flux of gravitational radiation to null infinity. The
``gauge transformations'' are now those coordinate transformations which leave this Bondi form invariant,
zeroes in the metric remain zeroes and $1/r$ terms get corrections of the same order (so that $m,c,\cb$ transform). 
Sending a pulse of
gravitational radiation or total energy $m(u_i)-m(u_f)$ to null infinity will change the functions $c,\cb$ by
a calculable amount. This corresponds to a change in the geodesic deviation of two objects at null
infinity. This is measurable by the gravitational memory effect so that one thus has measured the parameters
of a gauge transformation. 

What thus makes the gravitational memory effect physically significant is its direct connection to basic symmetries of GR. In view of this it is notable that the experimental prospects of measuring gravitational memory are quite promising. Even though a direct detection of memory from a single gravitational wave event by LIGO is unlikely (the memory signal $\sim 10^{-1}$ of the total gravitational wave strain), there exists the possibility of extracting the memory effect by statistical analysis from the cumulative data sourced by a collection of merger events. For instance, $\sim 90$ mergers similar to GW150914 yield an expected memory signal-to-noise ratio $\langle S/N_{tot} \rangle = 5$, whereas $\langle S/N_{tot} \rangle = 3$ is achieved by only $\sim 35$ events \cite{Lasky}. In light of the recent LIGO detections it seems probable that mergers of relevant size are relatively commonplace in the universe, making the expected frequency of future gravitational wave detections sufficient for measuring the memory in the coming years.

Having discussed the prospects of measuring the gravitational wave memory effect as well as the role of the gauge symmetries, let us finally contrast the situation with the U(1) memory. There is clearly some analogy between the two but also a considerable difference. In ED one has the overall U(1) gauge invariance and no need to define any new symmetries. In gravity the relevant transformations are a small carefully defined subset of general coordinate transformations and motivate a new symmetry transformation at null infinity, the BMS group \cite{sachs}.

\section{Color memory}
In Nature there are no massless free colored particles, asymptotic states have a mass
and do not propagate to null infinity. However, there is one context where one comes close: the
wave function of a large nucleus in an infinite momentum frame, probed with some
large scale phenomenon like deep inelastic scattering at large $Q^2$ and ``small $x$'',
$x=p^+/P^+$ = longitudinal momentum fraction of a parton in a nucleus. In infinite momentum 
frame the fast degrees of freedom are effectively frozen by time dilatation and can be
represented by a time independent color current. This sources classical color fields which describe dynamics
of soft and dense small $x$ degrees of freedom. Classical field description is justified by large
occupation numbers in $p,q$ phase space. However, classical fields are only an intermediate 
stage, in physics non-Abelian gauge theory is quantum theory and physics comes from an ensemble
average over classical fields sourced by an ensemble of sources.

\subsection{Color fields}
Let us first summarize the relevant color\footnote{Color conventions are
$D_\mu=\p_\mu-igA_\mu$, $A_\mu=A_\mu^a T_a$, $[T_a,T_b]=if_{abc}T_c$,
$T^a_{bc}=-if_{abc}$ for adjoint representation,
$F_\mn=i/g [D_\mu,D_\nu]=\p_\mu A_\nu-\p_\nu A_\mu-ig[A_\mu,A_\nu]$, under
a unitary gauge transformation $U(x):$ $A_\mu\to A_\mu^\prime=UA_\mu U^\dagger+i/g\, U\p_\mu U^\dagger$,
$F_\mn\to F_\mn^\prime=UF_\mn U^\dagger$.} fields \cite{blaizot}. 
We use the flat space light cone
coordinates with the mostly plus metric $ds^2=-2dx^+dx^-+dx^idx^i$, $x^\pm=(t\pm x)/\sqrt2$, 
$i=2,3$, $x_T^2=x^ix^i$, with the nucleus moving in the positive $x$ direction.  
In the limit of large $x\sim r$ the two transverse coordinates $x^i$ are
effectively the same as the S$^2$ angular coordinates $\theta_A$ (scaled by $r$).

The equation to be solved is
\be
D_\mu F^\mn = J^\nu=\delta^{\nu +}\rho(x^-,x^i)\ .
\la{YM}
\ee
Here $\rho$ is the color current of a nucleus moving in the $x$ direction in the infinite
momentum frame. It is crucial for the following that there is no $x^+$ dependence, there is
no time dependence due to infinite time dilatation. In contrast to the U(1) case, the formulation is not gauge invariant,
only gauge covariant. So we have to fix the gauge and the usual choice is the light cone gauge $A^-=-A_+=0$.
Then a current with only $+$ component and no $x^+$ dependence automatically satisfies
$D_\mu J^\mu=\p_+ J^+=0$, as required by \nr{YM}. 

However, $A^-=0$ is not yet complete gauge fixing and one can fix further either $A^i=0$ (Eq.\nr{ACOV})
or $A^+=0$ (Eq.\nr{LCgauge}). The former is called the 
covariant gauge (COV), since in it automatically $\p_\mu A^\mu=0$, and the latter, in unfortunate terminology, 
the light cone gauge (LC).
In both of these gauges $F^{-+}=F^{-i}=0$ while only $F^{+i}$ is non-zero,
$i=2,3$. 

One can formally avoid gauge fixing $A^-$ to zero by integrating $J^+(x^+,x^-,x^i)$ from the matrix equation
\be
D_+J^+=\p_+ J^+(x^+,x^-,x^i)-ig A^-(x^+,x^i)J^+(x^+,x^-,x^i)=0\ .
\la{A-}
\ee
This is clearly exceedingly complicated and anyway useless since for physical applications one also
has to include quantum fluctuations. This leads to an ensemble of color densities $\rho_a$, the distribution
of which is determined by a renormalisation group equation \cite{ilm}.

The covariant gauge (COV) corresponds to the gauge fixing
\be
A^\mu =(A^+(x^-,x^i),0,0,0),\quad A_\mu =(0,-A^+(x^-,x^i),0,0)\ ,
\la{ACOV}
\ee
for the vector potential. The absence of $x^+$ dependence means that 
$\p_\mu A^\mu=\p_+ A^+=0$ and the fact that $A^\mu$ has only one nonzero
component implies that the cross term in $F_\mn$ disappears. With the ansatz \nr{ACOV}
the field tensor, in the $(x^+,x^-,x^i)$ basis,  simply is
\be
F_\mn=\left( \begin{array}{ccc}0 & 0 & 0  \\ 0 & 0 & \p_i A^+(x^-,x^i) \\ 0&-\p_i A^+(x^-,x^i)&0 \end{array} \right)
\qquad ({\rm COV}).
\ee
The only non-zero component of the field tensor thus is $F_{-i}$ while $F_{+i}=0$. The latter 
implies that $F_{ti}=-F_{xi}$ so that, writing $F_{ti}=E_i,\,F_{ij}=-\epsilon_{ijk}B^k$,
\be
E_i=\epsilon_{xij}B_j\equiv \epsilon_{ij}B_j,\quad \epsilon_{23}=1
\ee
and
\be
F_{-i}=1/\sqrt2 \,(F_{ti}-F_{xi})=\sqrt2 E_i = \p_iA^+(x^-,x^i)\ .
\ee
Altogether we have $E_i=\epsilon_{ij}B^j$, $\epsilon_{23}=1$, $E_iB^i=0$, i.e., mutually orthogonal color
electric and magnetic fields, a good analogy for electromagnetic radiation. The relation of the fields to the
color current is obtained by solving $A^+$ from
\be
D_\mu F^{\mu +}=D_iF^{i+}=\p_i\p^i A^+=\p_i^2 A^+(x^-,x^i)=\rho(x^-,x^i)\ ,
\la{poisson}
\ee
i.e., by inverting the 2d transverse Poisson equation.

In the light cone gauge (LC) one asks for a potential of the form
\be
A^\mu=(0,0,A^i(x^-,x^j))\ . 
\la{LCgauge}
\ee
This is related to the previous by
transforming $A^+$ to zero in \nr{ACOV} by using the gauge transformation matrix 
\be
\p_- U^\dagger(x^-,x^i)=-igA^+(x^-,x^i)U^\dagger(x^-,x^i)\ ,
\ee
which is solved by the path ordered exponential
\be
U(x^-,x^i)=P\exp\biggl[ig\int_0^{x^-} dy^- A^+(y^-,x^i)\biggr]U(0,x^i)\ .
\la{U}
\ee
Because there is no $x^+$ dependence,  $A^-\to U\p_+ U^\dagger=0$, no $A^-$ is generated.
For the transverse components one has
\be
A^i(x^-,x^i)=i/g\,U\p_i U^\dagger\ .
\la{Ai}
\ee
The transverse potential thus is gauge equivalent to vacuum, the transverse field tensor vanishes:
\be
F_{ik}^a=\p_i A_k^a-\p_k A_i^a+gf_{abc} A_i^b A_k^c=0\ .
\la{trans}
\ee
Altogether the field tensor is
\be
F_\mn=\left( \begin{array}{ccc}0 & 0 & 0  \\ 0 & 0 & \p_- A^i(x^-,x^i) \\ 0&-\p_- A^i(x^-,x^i)&0 \end{array} \right)
\qquad ({\rm LC}).
\ee
The electric field is
\be
F_{-i}=\sqrt2\,E_i^{\rm LC}=\p_-A^i=i/g \p_- (U\p_i U^\dagger)=U\p_i A^+U^\dagger=\sqrt2 UE_i^{\rm COV}U^\dagger\ ,
\la{el}
\ee
as should. So we have symmetrically, depending on gauge, either $\sqrt2 E_i^{\rm COV}=\p_i A^+(x^-,x^i)$ 
or $\sqrt2 E_i^{\rm LC}=\p_- A^i(x^-,x^i)$.

\begin{figure}[!t]
\begin{center}
\includegraphics[width=0.8\textwidth]{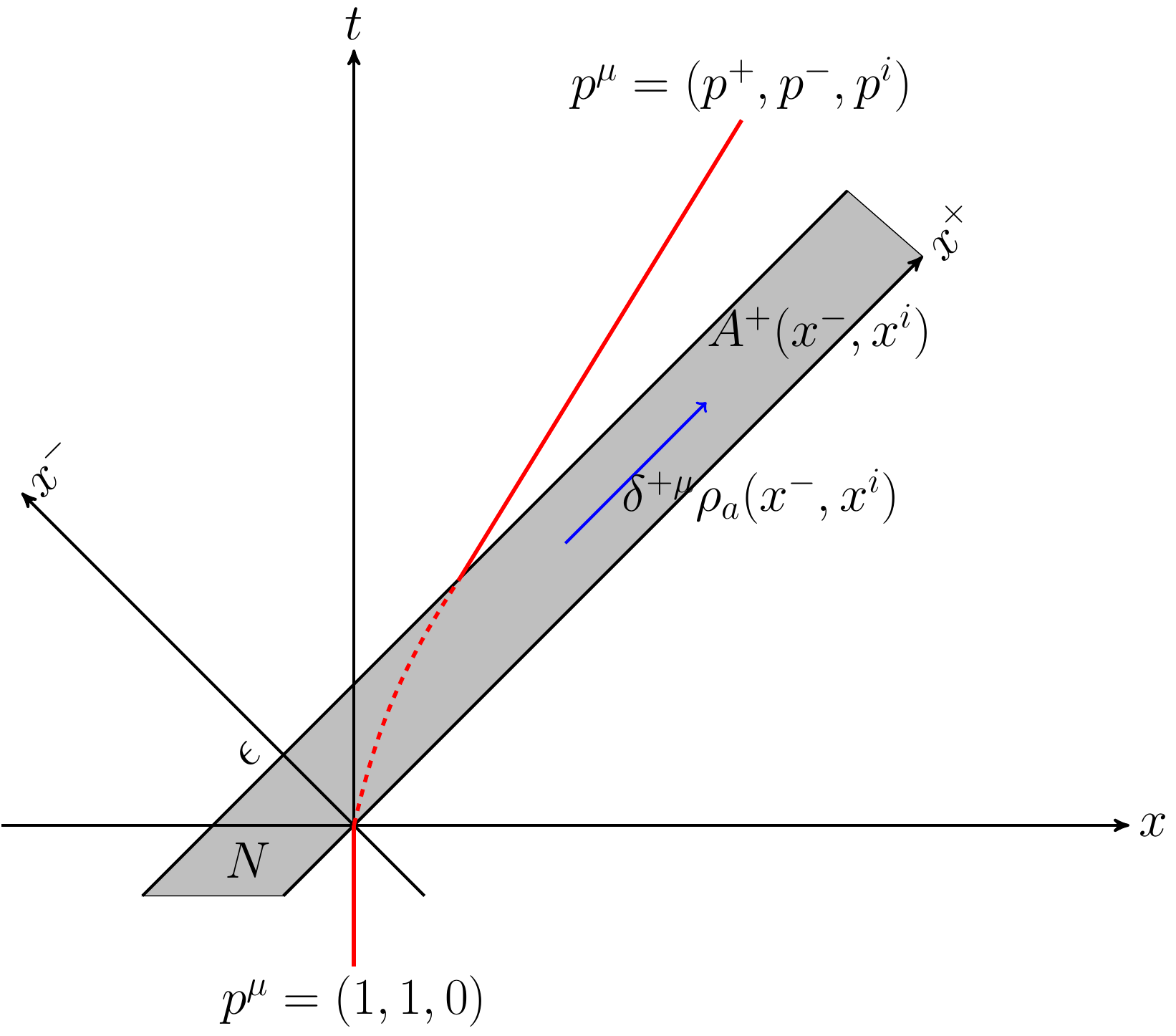}
\end{center}

\caption{\small  Interaction of a nuclear Yang-Mills field and a test quark. 
The nucleus $N$ is represented by a color current $J^+=\rho(x^-,x^i)$ and associated classical YM
field $A^+$. There is no dependence on the LC time $x^+$. 
The field extends over the range $0<x^-<\epsilon$ and $\epsilon\to0$ with increasing energy.  
In the transverse gauge $A^\mu=(0,0,A^i=i/g U\partial_i U^\dagger \theta(x^-))$,
fields in $0<x^-<\epsilon$ are given in \cite{blaizot}.
The collision with the test quark at rest 
accelerates the quark to transverse momentum $p^i$; this is the YM memory. Transverse coordinates are not shown
in the figure.
}
\la{burst}
\end{figure}

\subsection{Memory as a transverse kick}
We now have the classical color radiation fields -- in a fixed $A^-=0$ gauge -- and the next task is to formulate the analogue
of the memory equation \nr{memory}, i.e., how a transverse momentum kick of a test quark can be computed. 
For this we need a generalisation of the Lorentz force and
the equations of motion of a colored test particle in a known color field, the 
Wong equations \cite{wong,manuel,venuwong}. The same equations have recently been used \cite{Kajantie:2019hft} in a study of
collisions of a nucleus with a static test quark, the emphasis there was on the gluon radiation caused by the acceleration
of the quark.

Generalising the electrodynamic action for a point particle following the path $x^\mu=x^\mu(\tau)$ 
by introducing a color vector $Q^a(\tau)$, the Wong equations can be derived from the action
\ba
S&=&\int d\tau\bigl[-m{ds\over d\tau} +g {\dot x}^\mu(\tau) A_\mu^a(x^\alpha(\tau))\, Q_a(\tau)\bigr]
=\int d\tau L[x^\mu(\tau),\dot x^\mu(\tau)]\nn
&=&-m\int ds+\int d^4x\biggl[\int d\tau\,\delta^4(x^\mu-x^\mu(\tau))gQ_a(x)u^\mu\biggr]A_\mu^a(x)\ .
\la{S}
\ea
For a given vector potential $A_\mu^a$ the equations are extremal equations for a particle path
$x^\mu=x^\mu(\tau)$. Defining first
\be
p^\mu=mu^\mu=m{dx^\mu\over d\tau}
\ee
they are ($Q\cdot F\equiv Q_aF^a =2\Tr QF$)
\be
{dp^\mu\over d\tau}=gQ\cdot F^{\mu\nu}{dx_\nu\over d\tau}, 
\quad {dQ^a\over d\tau}=-gf_{abc}u^\mu A^b_\mu Q^c\ .
\la{eom}
\ee
Here the proper time dependence of $Q(\tau)$ follows elegantly from demanding that the
extremal equations deried from the Lagrangian in \nr{S}
give the correct non-Abelian cross term in $F^\mn$ in the first equation. 
Note that the equation for $p_\mu$ explicitly conserves the mass shell condition $p_\mu p^\mu=-m^2$.

The equation for $\dot Q\equiv dQ/d\tau$ also follows from the conservation law $D_\mu J^\mu=0$ for the current
\be
J^\mu=\int d\tau Q(\tau) u^\mu(\tau)\delta^4(x-x(\tau))\ .
\ee
In matrix form ($u^\mu\p_\mu=\p_\tau$)
\be
\dot Q-igu^\mu A_\mu Q=u^\mu(\p_\mu-igA_\mu)Q =u^\mu D_\mu Q=0\ .
\ee
This first order matrix equation can be integrated to give
\be
Q(\tau)=P\exp\biggl[ig\int_0^\tau dx^\mu A_\mu(x)\biggr]Q(0)\ .
\la{Qsol}
\ee 

Consider then the Lorentz force equation for $\mu=-,i,+$.
For $\mu=-$ one simply has $F^{-+}=F^{-i}=0$ and
\be
{dp^-\over d\tau}=0\quad \Rightarrow\quad {p^-\over m}=u^-={dx^-\over d\tau}={\rm constant} 
\quad\Rightarrow\quad x^-(\tau)=u^-\tau\ .
\ee
The fact that the $\mu=-$ component is so simple basically follows from the time or $x^+$ independence of the gluon radiation
burst in \nr{YM}. The Lorentz force equation trivially conserves $p^2=-2p^+p^-+p_i^2=-m^2$ from which
\be
p^-{dp^+\over d\tau}=p_i{dp^i\over d\tau}\ .
\la{mshell}
\ee
Thus only the equation for $p^i$  is needed. In the $A^-=0$ gauge,
\ba
{dp^i(\tau)\over d\tau}=m{d^2x^i(\tau)\over d\tau^2}&=& gQ\cdot F^{i+}{dx_+\over d\tau}=-gu^-Q(\tau)\cdot \p_- A^i(x^-,x^k(\tau))\nn
&=&-gu^-\sqrt2\,Q(\tau) \cdot E^i(x^-,x^k(\tau))\ ,
\la{pi}
\ea
where we inserted $x_+=-x^-$ and remember that  $x^-=u^-\tau$.
Eq.\nr{pi} with the color electric field is obviously the analogue of the 
simple equation $\ddot x=eE_x$  in the Abelian ED case \nr{memory}.

Together with Eq.\nr{eom} for $Q(\tau)$,  Eq.\nr{pi}
is a very complicated 2nd order differential equation for the transverse
coordinate $x^i(\tau),\,\,i=2,3$. First, we are given a color distribution $\rho_a(x^-,x^i)$ in the
infinite momentum wave function of a nucleus, one may imagine a Gaussian in all variables.
It serves as the inhomogeneous source term of a 2d Poisson equation \nr{poisson} for the
potential $A^+(x^-,x^i)$. Using this one can define the path ordered exponential
$U(x^-,x^i)$ in Eq.\nr{U} which, via Eq.\nr{Ai} gives the transverse vector potential $A^i(x^-,x^i)$.
When this background field is given one can solve the rotation of $Q(\tau)$ from \nr{Qsol}. All these equations
depend implicitly on the quantity to be solved, $x^i(\tau)$. Assuming the test quark is initially at rest,
$x^i(0)=x^i_0,\quad x^{i\prime}(0)=0$, one can, in principle,  solve $x^i(\tau)$ and $p^i(\tau)$.
Yang-Mills memory then is simply given by the total transverse kick,
\be
\Delta p^i = p^i(\tau_f)-p^i(0)\ .
\la{mem}
\ee

The above computation was carried out in the $A^-=0$ gauge. 
Transformation within the two gauges in this class is
\be
F_{\rm COV}^{i+}=U^\dagger F_{\rm LC}^{i+}U=\p_iA^+_{\rm COV}=U^\dagger \p_-A_i^{\rm LC}U,\qquad
Q_{\rm LC}=UQ_{\rm COV}U^\dagger\ .
\ee
and the result, which is $\sim\Tr QF^{i+}$, is explicitly invariant and physical under these transformations. 
Restoring $A^-$ as in \nr{A-} is also possible, but does not change the fact that memory is defined in a fixed
gauge. In ED, in the formulation of \cite{garfinklegrav}, the gauge transformations $U$ simply disappear from
the definition.

\section{Simplification in quantum theory}
To integrate \nr{pi} and to use the remaining piece of information, the $\tau$ or $x^-$ derivative of $Q$ in \nr{eom},
we manipulate \nr{pi} as follows: 
\ba
{dp^i(x^-)\over dx^-}&=&-g[\p_-(Q_a(x^-)A^i_a)-A^i_a\p_-Q_a]\nn
&=&-g\left[{d(Q_a(x^-)A^i_a)\over dx^-}+g f_{abc}A^a_iA^b_k\,Q_c(x^-) {d x^k\over dx^-}\right]\ .
\la{pif}
\ea
In the present gauge $A_\pm=0$ and in the sum in \nr{eom} only the spatial term $u^kA_k^b$ remains. In \nr{pif} we could
replace $gf_{abc}A_i^aA_k^b=-\p_iA_k^c+\p_kA_i^c$ (since $F_{ik}=0$), but this complicated term does not vanish.

In quantum theory of Color Glass Condensate (CGC) the situation is actually much simpler. There one does not compute the
fields for a fixed color distribution $\rho_a$ in \nr{YM}, but integrates over a distribution thereof in order to compute
expectation values, see, e.g., \cite{ilm} Section 2. These are diagonal in color. On the average, color density
vanishes, $\langle\rho_a\rangle=0$ and the starting point is the charge density correlator
\be\label{rhorho}
\langle \rho_a(x^-,x^i)\,\rho_b(y^-,y^j)\rangle=\delta_{ab}\delta(x^--y^-)\delta^{(2)}(x-y)\lambda_A(x^-)\ ,
\ee
where $\lambda_A(x^-)$ is the average color charge squared of valence quarks per color and per volume
(so that its integral over $x^-$ is the average transverse density). 
From this one can compute $\langle A_a^+(x^-,x^i)\,A_b^+(y^-,y^j)\rangle$ and further correlators of 
the type $ \langle A^a_iA_j^b\rangle$.
These are all diagonal, $\sim\delta_{ab}$. When applied to \nr{pif},  the last term in it vanishes, due to the antisymmetry 
 of $f_{abc}$. Thus we can immediately integrate and find
\be
p^i(x^-)=m{dx^i\over d\tau}=-gQ_a(x^-)\, A^i_a(x^-,x^i(x^-))\ .
\la{mem1}
\ee
We already have solved $x^-(\tau)=u^-\tau$ and the transverse coordinate $x^i=x^i(\tau)$
can also be solved from here. Finally, the mass shell condition \nr{mshell}
gives, after integration over $\tau$, 
\be
p^+(\tau)=m{dx^+\over d\tau}={(gQ\cdot A^i)^2\over 2p_-}+{\rm const}={p_i^2(\tau)+m^2\over 2p^-}\ .
\ee

The above is a great simplification relative to \nr{pif}, but in full quantum theory the coordinates disappear 
and what matters is the expectation value \cite{iancu}
\be
p_T^2=\langle p^ip^i\rangle = g^2 Q_bQ_b \langle A_a^i A_a^i\rangle\ .
\la{memfin}
\ee
The signal of color memory is thus $p_T\sim gQ |A^i|$, the magnitude of the transverse color field 
in the gauge $A^-=0$, generated by the passing of the nucleus. This generalises the memory in ED,
$\Delta p_B=q(A_B(u_f)-A_B(u_i))$ in the gauge $A_u=0$, derived in Eqs. \nr{memory} and \nr{MADeltaA}.
Deriving the physical magnitude of the kick is one of the achievements of the theory of CGC, especially for
very large nuclei. A dense system of gluons saturates and generates a dynamical large scale, the saturation scale
$Q_s\sim $ few GeV. The physical magnitude of the kick then is, for dimensional reasons, $\sim Q_s$.

Since \nr{memfin} is the main result of this article, it might be useful to add some detail on how its magnitude is
computed, in the simplest possible way. One assumes that the nucleus is infinitely Lorentz contracted, $x^-$ dependence
is $\delta(x^-)$, and that 
the color density of the nucleus fluctuates according to the Gaussian distribution
\be
W[\rho_a({\bf x})] = \exp\left[-\int d^2z{1\over 2\lambda}(\rho_a({\bf z}))^2\right]\ ,
\label{W}
\ee
where only dependence on the transverse coordinate ${\bf z}$  is needed. Average charge then vanishes and quadratic correlators
are given by \nr{rhorho} with $x^-$ dependence removed (only $x^-=0$ contributes). Physics, properties of the nuclear wave function,
is embedded in the constant $\lambda$. According to \nr{Ai} $A_i({\bf x})$ is simply 
related to the Wilson line \nr{U} and the expectation value of the correlator becomes
\be
\langle{\rm Tr}A_i({\bf x})A_j({\bf y})\rangle_\rho=\fra12\langle A_i^aA_j^a\rangle_\rho=\fra1{g^2}\langle{\rm Tr}
(U\p_iU^\dagger({\bf x})\p_jU\,U^\dagger({\bf y}))\rangle_\rho \ .
\ee
We need the magnitude of $A_i$ at a point, so ${\bf x}\to {\bf y}$, $U^\dagger U=1$ and one has to compute in this limit
\be
\langle{\rm Tr}A_i({\bf x})A_j({\bf y})\rangle_\rho=\fra1{g^2}\p_i^x\p_j^y\langle{\rm Tr}
(U^\dagger({\bf x})U({\bf y}))\rangle_\rho\ .
\label{pipj}
\ee
$U$ is $\sim\exp(-igA^+)$ by \nr{U} and $A^+=(1/\p_i^2)\rho$ by the Poisson equation \nr{poisson} so that the product
$UU^\dagger$ is exponential of a linear functional of $\rho$ and the Gaussian integral over the weight function \nr{W}
can be carried out. The 2d Poisson equation has a logarithmic divergence, though, and cutting this away by $1/k_T^2\to 1/(k_T^2+m^2)$,
the result is
\be
\langle{\rm Tr}(U^\dagger({\bf x})U({\bf y})\rangle_\rho=\exp\left[-{\lambda g^2N_c\over 4\pi m^2}(1-mr_TK_1(mr_T))\right]
\approx \exp\left[-Q_s^2\,{r_T^2\over 4\pi}\log{1\over r_T\Lambda}\right]\ ,
\ee
where $Q_s^2=\fra12\lambda g^2N_c$, $r_T=|{\bf x}-{\bf y}|$, $\Lambda=\fra12 e^{\gamma_E-\fra12}m$ and we have taken the limit $m\to0$.
The final step of taking the derivatives in \nr{pipj} leads to
\be
\lim_{{\bf x}\to{\bf y}}\langle A_i^a({\bf x})A_i^a({\bf y})\rangle = Q_s^2\fra1{g^2\pi}\lim_{r_T\to0}\log{1\over r_T\Lambda}\ ,
\ee
a result proportional to $Q_s^2$ but now divergent at small distances. This reflects the small size of the probe, a single quark, the kick of which
one is studying. In momentum space this would correspond to the tail at large momentum. Dynamics at small distances or 
at large momenta has to cut-off the divergence, but the above outline of a computation is offered here as an illustration
of the difficulties testing Yang-Mills memory in a nuclear environment has to face. 

It is interesting to ask if the color memory effect could be experimentally verified. The observable linked most naturally with the kick \nr{memfin} is the color dipole cross-section \cite{ymmemo2} and the prospects for its extraction in the future Electron-Ion Collider is discussed in \cite{Aschenauer:2017jsk}. The detection seems a plausible scenario, albeit in a highly convoluted environment.

Finally, let us point out that in principle, a color rotation of the quark could also be a signal of the memory \cite{pate}. However, in
\cite{Kajantie:2019hft} it was shown that in the relevant limit of very high energies, $\epsilon\to0$ in Fig.~\ref{burst},
this rotation vanishes.

\section{Conclusions}

We have discussed the memory effect in Abelian and non-Abelian gauge theories. In Abelian electrodynamics the discussion
can be entirely formulated using gauge invariant quantities, electric and magnetic fields \cite{bieriED}. The effect manifests itself as a change
in the transverse momentum of a test charge caused by the passing of the radiation pulse. It has two components, an
ordinary kick caused by the change in the radial component of the electric field and a null kick caused by flux of charge
to null infinity. The latter, more interesting, part requires that there be massless charged particles since massive ones
do not get to null infinity. Such ones do not exist but could exist so this is a useful conceptual exercise as an analogy to
gravity, there certainly massless particles carry energy to null infinity.

The U(1) memory could be analysed entirely in terms of gauge invariant variables. However, one can also fix the gauge
at null infinity and one thus has a means of determining physically a magnitude of a gauge transformation \cite{winicour}.
This is a ``large'' (non-zero) gauge transformation at asymptotic infinity. One can
formulate this as a symmetry of ED at infinity, but it is essentially just U(1) gauge invariance.

In the non-Abelian case there in nature are no classical fields of the type used in the discussion of U(1) memory, QCD
is in the confinement phase, gluons dress themselves to glueballs and do not propagate to null infinity. However,
classical YM fields supplemented with their quantum fluctuations are ubiquitous in discussions of dynamics of large nuclei, 
in certain experimental conditions they become very dense systems with large occupation numbers so that use of classical
fields is motivated. Large transverse densities produce effectively a large energy scale, saturation scale, so that the coupling 
becomes - optimistically - weak. In this framework, the theory of Color Glass Condensate, 
one can immediately find a generalisation of the key
U(1) memory equation \nr{MADeltaA}, in which a physical quantity, transverse kick, is expressed as color times gauge potential
in a certain gauge. Initially one has a linear equation of the type  in \nr{mem1}, but its expectation value vanishes and 
what matters is the quantum expectation value of its square in \nr{memfin}.

The discussion of memory in YM theory is thus not manifestly gauge invariant and physical results are obtained only after
gauge fixing in quantum theory. It is thus not useful as an analogue model of gravitational radiation. Rather the other way
round, it can be used to reinterpret some well known properties of the theory of CGC.

\vspace{0.8cm}
 {\bf Acknowledgements} Miika Sarkkinen is supported by the Finnish Cultural Foundation. We thank L.~McLerran,  T.~Lappi, and R.~Paatelainen for discussions on Color Glass Condensate and L.~Bieri, D.~Garfinkle, D.~Nichols, and B.~Oblak for discussions on gravitational radiation memory, in particular at a Solvay workshop in May 2018.

\end{document}